# Bright X-ray Flares in Gamma-Ray Burst Afterglows

D. N. Burrows[1*], P. Romano[2], A. Falcone[1], S. Kobayashi[1,3], B. Zhang[4], A. Moretti[2],

P. T. O'Brien[5], M. R. Goad[5], S. Campana[2], K. L. Page[5], L. Angelini[6,7], S. Barthelmy[6],

A. P. Beardmore[5], M. Capalbi[8], G. Chincarini[2,9], J. Cummings[6], G. Cusumano[10],

D. Fox[11], P. Giommi[8], J. E. Hill[1], J. A. Kennea[1], H. Krimm[6], V. Mangano[10],

F. Marshall[6], P. Mészáros[1], D. C. Morris[1], J. A. Nousek[1], J. P. Osborne[5],

C. Pagani[1,2], M. Perri[8], G. Tagliaferri[2], A. A. Wells[5], S. Woosley[12], N. Gehrels[6]

---

[1] Department of Astronomy & Astrophysics, Pennsylvania State University, 525 Davey Lab, University Park, PA, 16802, USA

[*] To whom correspondence should be addressed. E-mail: burrows@astro.psu.edu.

[2] INAF-Osservatorio Astronomico di Brera, Via Bianchi 46, 23807 Merate, Italy

[3] Center for Gravitational Wave Physics, Pennsylvania State University, 104 Davey Lab, University Park, PA, 16802, USA

[4] Department of Physics, University of Nevada, Box 454002, Las Vegas, NV, 89154-4002, USA

[5] Department of Physics and Astronomy, University of Leicester, University Road, Leicester LE1 7RH, UK

[6] NASA/Goddard Space Flight Center, Greenbelt, MD, 20771, USA

[7] Department of Physics and Astronomy, Johns Hopkins University, 3400 North Charles Street, Baltimore, MD, 21218, USA

[8] ASI Science Data Center, Via Galileo Galilei, 00044 Frascati, Italy

[9] Dipartimento di Fisica, Università degli studi di Milano-Bicocca, Piazza delle Scienze 3, 20126, Milan, Italy

[10] INAF- Istituto di Astrofisica Spaziale e Fisica Cosmica Sezione di Palermo, Via Ugo La Malfa 153, 90146 Palermo, Italy

[11] Department of Astronomy, California Institute of Technology, MS 105-24, Pasadena, CA, 91125, USA

[12] Department of Astronomy & Astrophysics, University of California, Santa Cruz, CA, 95064, USA

Gamma-ray burst (GRB) afterglows have provided important clues to the nature of these massive explosive events, providing direct information on the nearby environment and indirect information on the central engine that powers the burst. We report the discovery of two bright X-ray flares in GRB afterglows, including a giant flare comparable in total energy to the burst itself, each peaking minutes after the burst. These strong, rapid X-ray flares imply that the central engines of the bursts have long periods of activity, with strong internal shocks continuing for hundreds of seconds after the gamma-ray emission has ended.

Gamma-ray bursts (GRBs) are the most powerful explosions since the Big Bang, with typical energies around $10^{51}$ ergs. Long GRBs (duration > 2 s) are thought to signal the creation of black holes by the collapse of massive stars ($1-4$). The detected signals from the resulting highly relativistic fireball consist of prompt gamma-ray emission (from internal shocks in the fireball) lasting for several seconds to minutes, followed by afterglow emission (from external shocks as the fireball encounters surrounding material) covering a broad range of frequencies from radio through X-rays ($5-7$). Because of the time needed to accurately determine the GRB position, most afterglow measurements have been made hours after the burst, and little is known about the characteristics of afterglows in the minutes following a burst, when the afterglow emission is actively responding to inhomogeneities in both the fireball and the circumburst environment.

The *Swift* (*8*) X-ray Telescope (XRT)(*9*) provides unique and novel X-ray observations of young Gamma-ray Burst (GRB) and X-ray Flash (XRF) afterglows, beginning in the first few minutes after the burst. (Here we use the terms "burst" and "prompt emission" to refer to the burst seen in hard X-rays and gamma rays, and the term "afterglow" to refer to the soft X-ray, optical, and radio emission seen after the end of the detectable hard X-ray prompt emission.) Between 23 December 2004 and 5 May 2005, the XRT observed 13 afterglows within 200 seconds of the onset of GRBs discovered by the *Swift* Burst Alert Telescope (BAT)(*10*). In most cases the XRT found a bright, monotonically decaying afterglow (*11 – 14*). By contrast, the afterglows of both XRF 050406 and GRB 050502B (*15*) were a factor of 10 – 1000 times fainter than previous XRT-detected afterglows at T+100 s, but brightened rapidly several minutes later before decaying back to their pre-flare fluxes (Fig. 1). The afterglow of XRF 050406 brightened by a factor of 6 between 150 and 213 s post-burst and is similar to X-ray flares observed in a few previous cases (*16*). GRB 050502B is qualitatively different, with a giant flare that brightened by a factor of ~500 to a peak at T+740 s. This flare contained roughly as much energy (~$9 \times 10^{-7}$ ergs cm$^{-2}$, 0.3-10 keV) as the prompt emission observed by the BAT ($8 \times 10^{-7}$ ergs cm$^{-2}$, 15-350 keV), something never before seen and quite unexpected.

The rise and fall of the flare in XRF 050406 are both very steep. Following standard practice, we characterize the X-ray afterglow decays as power-laws with the X-ray flux varying as $F_x \propto t^\alpha$, where $t$ is the time since GRB onset. Using this same form to describe the X-ray flare, we find $\alpha = +4.9 \pm 0.3$ during the rising portion, and $\alpha = -5.7 \pm 0.6$ during the decay, with $\delta t/t_{peak} \sim 0.3$ and 0.6 for the rise and fall times, respectively. (The flare slopes are more symmetrical if the underlying power-law afterglow decay is subtracted off.) Such large slopes cannot be explained by external forward shocks, where the radiation physics implies a slower decay, with the decay time $\delta t$ comparable to the post-shock time $t$ (*17*). The shape of the flare is reminiscent of that expected for an external reverse shock, created in the outflow when the forward shock is slowed significantly by interaction with an external medium. However, reverse shocks are expected to be far less steep and should be seen in the optical, not the X-ray, band. Synchrotron self-Compton (SSC) models may be able to produce X-ray emission from a reverse shock, but only for carefully balanced conditions (*18*). A far more natural explanation for this flare is continuation of strong internal shocks to a time of at least T+213 s.

The flare in GRB 050502B is slower, with $\delta t_{decay}/t \sim 1$, but the sharp spike at T+740 s (seen in the hard band in Figure 2b) argues against an external shock mechanism (*17*). If produced by internal shocks, energy production by the central engine must continue for at least 740 s after the burst begins in this case.

Extended activity in the central engine can explain both of these flares. The central engine becomes active again around 150 seconds and 300 seconds after the burst for XRF 050406 and GRB 050502B, respectively. The duration of the flare directly measures the duration of the central engine activity because the observed time sequence essentially follows the central engine time sequence (*19*).

With the exception of the flare at 213s, the count rate of XRF 050406 was very low and detailed time-resolved spectroscopy is not possible. We can obtain some information on the spectral evolution of the flare, however, by dividing the data into two energy bands and examining light curves in those bands. Figure 2a shows the light curves in the 0.2-0.7 keV and 0.7-10 keV bands, together with the ratio of these bands. There is significant spectral evolution during the first 400 seconds of this afterglow. During the rising flare (about T+180s to T+200s) the hard band flux spikes up rapidly while the soft band flux remains constant, indicating that the flare is harder spectrally than the underlying afterglow. In fact, the rising portion of the flare contributes no significant flux between 0.2 and 0.7 keV, while increasing the count rate in the 0.7-10 keV band by a factor of four. This provides strong constraints on the flare mechanism: it is difficult to quadruple the flux in the high energy band while the low energy band remains constant, unless the flare is strongly absorbed by a column of neutral gas (with $N_H \sim 10^{21}$ cm$^{-2}$) that does not affect the underlying afterglow. However, the soft band peaks during the time bin following the overall peak, indicating that the emission softens substantially as the flare decays away. If absorption is invoked for the rising portion of the flare, then the absorption seems to decrease significantly during the flare decay, suggesting that the absorbing gas may be ionized by the flare (*20,21*). Following the flare the band ratio returns to a value consistent with the pre-flare values. A band ratio plot of GRB 050502B also shows clear indications of spectral variations (Figure 2b), with a trend similar to that in XRF 050406 (hardening at the beginning of the flare and gradually softening as the flare progresses), although in this case both bands increase during the rising portion of the flare.

We have referred to these events as X-ray flares because they were not detected by the higher-energy BAT instrument on *Swift*. This is presumably due to the higher sensitivity of the XRT and to the steep spectral energy indices of the afterglows ($\beta = -1.3$ for the flare in XRF 050406 and $\beta = -1.4$ for the flare in GRB 050502B, where $F_x \propto E^\beta$). In the case of XRF 050406, the prompt emission was classified as an X-ray Flash due to its relatively soft spectrum. The discovery of these large X-ray flares in the afterglows raises the possibility that these flares themselves would be classified as XRFs, had they not been preceded by the higher-energy bursts detected by the BAT. In fact, GRB 050502B appears to be a remarkable combination of a normal GRB followed more than 10 minutes later by an XRF of comparable fluence. [There is indirect evidence for a similar sequence in GRB 031203 (*22*)]. If a normal, relatively hard burst like GRB

050502B can produce such a bright X-ray flare through late-time internal shocks, this suggests that XRFs themselves may be related to the characteristics of the central engine rather than to geometrical effects (*23,24*). However, we note that the long durations and smooth temporal profile observed in these X-ray flares are quite different from those typically detected in XRFs (*25*).

Both of these afterglows appear to be dominated by long periods of energy production by the central engine, leading to a long period of X-ray emission from internal shocks extending long past the cessation of gamma-ray production. Such activity has been previously suggested as an explanation for extended GRB tails observed by the *BATSE* instrument (*26*). In addition to the large flares several minutes after the burst, the plateaus or bumps beginning several hours later imply either that significant energy is still being injected into the blast wave by refreshed shocks (*27*), that the external shocks are encountering dense clumps in the nearby interstellar medium (*28*), or that the internal shocks are still continuing up to several days after the burst in the observer's frame. The latter possibility would require that the central engine operated for a time-scale of days, possibly due to fallback of material into the central black hole (*29,30*).

If the central engine is still pumping significant energy into the blast wave at such late times, how can one explain the short duration of the gamma-ray emission? The late internal shocks that produce the flares must produce lower energy photons than the earlier internal shocks. This can be explained by higher bulk Lorentz factors, which result in lower magnetic fields at the larger radius reached by the internal shocks at these late times. For the internal shock model, the typical peak energy is
$E_p \propto L^{1/2} R^{-1} \propto L^{1/2} \Gamma^{-2} \delta t^{-1}$ (*31*), where $L$ is the luminosity at the flare epoch, $R$ is the internal shock distance, $\Gamma$ is the Lorentz factor, and $\delta t$ is the variability time-scale. For a radius $R$ that is 3-10 times larger for the flare than the typical internal shock radius of the burst, $E_p$ may be 10 times lower. This is consistent with the detection of these flares in the X-ray band rather than in gamma-rays, and with the large δt observed in the X-ray flares. The higher bulk Lorentz factors may arise because these late-time internal shocks benefit from a low-density channel through the progenitor star, previously excavated by the jets that produce the original gamma-ray burst. These later outflows would then have lower amounts of entrained baryons and higher Lorentz factors, leading to lower photon energies at later times and pushing the late-time internal shock emission below the BAT energy band (15-150 keV).

**Table 1.** (Supplemental table in published version.) Properties of the bursts presented here. XRF 050406 was discovered by the *Swift* BAT instrument at 15:58:48 UTC on 6 April 2005. It had a very soft spectrum (power-law spectrum $N(E) \propto E^{-\Gamma}$ with photon index $\Gamma=2.4$) and is classified as an X-ray Flash (XRF; *25*). The *Swift* observatory performed a prompt slew to the burst location, pointing the XRT and the UV-Optical Telescope (UVOT) toward the burst in ~84 seconds and the XRT executed its normal sequence of readout modes (*32*). No bright source was found in the first 2.5 s XRT exposure, but the count rate began climbing rapidly about 180 s after the BAT trigger (Figure 1a). GRB 050502B was a typical multi-peaked burst discovered by the *Swift* BAT instrument at 09:25:40 UTC on 2 May 2005. The *Swift* observatory performed a prompt slew and the XRT began collecting data ~63 seconds after the burst trigger. No bright source was found in the first 2.5 s exposure. About 300 seconds after the burst, the X-ray intensity began to rise steeply, switching the XRT from Photon Counting (PC) mode into Windowed Timing (WT) mode through the peak at 740 s post-burst (Figure 1b; see ref. *32* for a discussion of XRT readout modes)

| Burst Name | Trigger time | $T_{90}$ (s)[†] | Photon Index | Prompt Fluence (ergs cm$^{-2}$) | Ref. |
|---|---|---|---|---|---|
| XRF 050406 | 6 April 2005 15:58:48 UTC | 5 ± 1 | 2.4 ± 0.3 | 9 x 10$^{-8}$ | 33, 34 |
| GRB 050502B | 2 May 2005 09:25:40 UTC | 17.5 ± 0.2 | 1.6 ± 0.1 | 8 × 10$^{-7}$ | 35, 36 |

[†] $T_{90}$ is the burst duration, defined as the time within which 90% of the photons arrived.

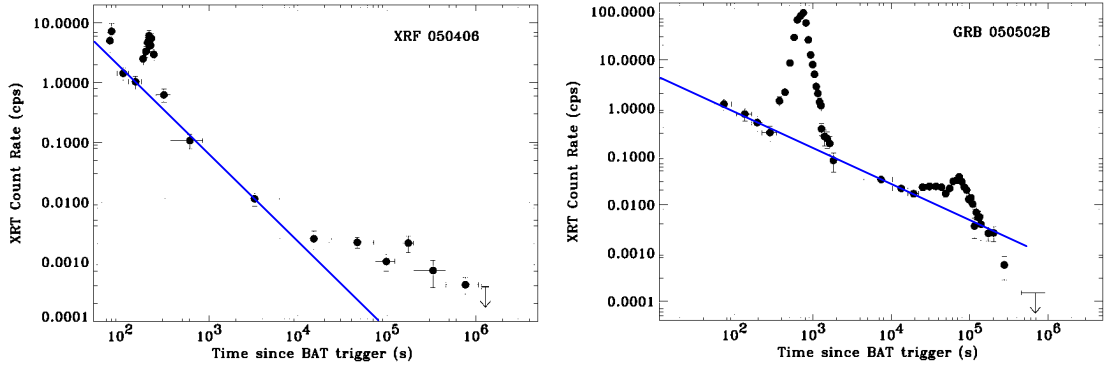

**Fig. 1.** Background-subtracted X-ray light curves of the afterglows of XRF 050406 (**A**) and GRB 050502B (**B**). For XRF 050406 we obtained a total exposure time of 155 ks distributed over 17 days; for GRB 050502B the total exposure time was 176 ks over 11 days. The solid lines represent power-law fits to the underlying afterglow decays from about 100 s to 10,000 s (power-law index is -1.5 ± 0.1 and -0.8 ± 0.2 for XRF 050406 and GRB 050502B, respectively). The bright X-ray flares are superposed on this underlying power-law decay. At later times the XRF 050406 light curve flattens, while the GRB 050502B has several bumps, both suggesting late-time energy injection into the external shock or continued internal shock activity. The rapid decline in count rate for GRB 050502B at T> $10^5$ s indicates a possible jet break at around 1-2 days post-burst.

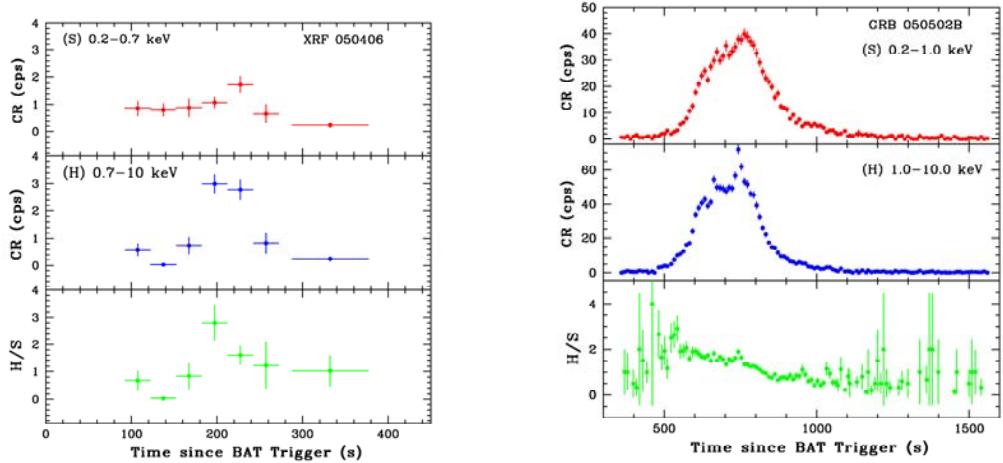

**Fig. 2.** Band ratio plots, showing spectral variations during the flares. In both cases the top two panels show the count rates in the soft and hard bands, while the bottom panel shows the band ratio (hard/soft). The boundaries between the hard and soft bands were chosen independently for each burst to provide comparable numbers of counts in the two bands. **(A)** Light curves in 0.2-0.7 and 0.7-10 keV bands for XRF 050406. The rising part of the flare is significantly harder than the decaying part of the flare or the underlying afterglow, suggesting that the flare is not caused by the external shock responsible for the underlying afterglow. **(B)** Light curves in 0.2-1.0 and 1-10 keV bands for GRB 050502B. The band ratio shows strong spectral variations during the large flare, with the ratio of counts in these bands decreasing by a factor of 4. The sharp spike in the hard band at about 740 s supports the internal shock interpretation for this flare, since such sharp features cannot easily be produced by external shocks.

[37] This letter is based on observations with the NASA Swift gamma-ray burst Explorer. We thank the Swift operations team for their support. The authors acknowledge support from NASA in the US, ASI in Italy, and PPARC in the UK.